\documentclass[twocolumn,aps,showpacs,nofootinbib,superscriptaddress,prd]{revtex4}
\usepackage{mathrsfs}
\usepackage{epsfig}
\usepackage{textcomp}
\usepackage{amsmath}
\usepackage{graphicx}
\usepackage{amssymb}
\usepackage{tikz}
\usepackage{pgfplots}
\usepackage{pgf-umlsd}
\usepackage{ifthen}
\usepackage{hhline}

\begin{document}

\title{Improving the measurement of the Higgs boson-gluon coupling using convolutional neural networks at $e^+e^-$ colliders}

\author{Gexing Li}
\email{ligx@ihep.ac.cn}
\affiliation{Institute of High Energy Physics, Chinese Academy of Sciences, Beijing 100049, China}
\affiliation{School of Physics Sciences, University of Chinese Academy of Sciences, Beijing 100039, China}

\author{Zhao Li}
\email{zhaoli@ihep.ac.cn}
\affiliation{Institute of High Energy Physics, Chinese Academy of Sciences, Beijing 100049, China}
\affiliation{School of Physics Sciences, University of Chinese Academy of Sciences, Beijing 100039, China}

\author{Yan Wang}
\email{wangyan728@ihep.ac.cn}
\affiliation{College of Physics and Electronic Information, Inner Mongolia Normal University, Hohhot 010022, China}
\affiliation{Institute of High Energy Physics, Chinese Academy of Sciences, Beijing 100049, China}
\author{Yefan Wang}
\email{wangyefan@ihep.ac.cn}
\affiliation{Institute of High Energy Physics, Chinese Academy of Sciences, Beijing 100049, China}
\affiliation{School of Physics Sciences, University of Chinese Academy of Sciences, Beijing 100039, China}


\begin{abstract}
In this paper we propose to use convolutional neural networks (CNNs) to improve the precision measurement of
the Higgs boson-gluon effective coupling at lepton colliders.
The CNN is employed to recognize the Higgs boson and a $Z$ boson associated production process,
with the Higgs boson decaying to a gluon pair and the $Z$ boson decaying to a lepton pair
at the center-of-mass energy 250 GeV and integrated luminosity 5 ab$^{-1}$.
By using CNNs, the uncertainty of the effective coupling measurement can be decreased from $1.94\%$ to about $1.28\%$ using the PYTHIA data
and from $1.82\%$ to about $1.22\%$ using the HERWIG data in the Monte Carlo simulation.
Moreover, the performance of CNNs using different final state constituents shows
that the energy distributions of the leading and subleading jets constituents play a major role in the identification
and the optimal uncertainty of effective coupling using CNNs is reduced by about $35\%$ compared to that using conventional method.
\end{abstract}

\maketitle

\section{Introduction}

The Higgs boson occupies a distinct place in the Standard Model (SM) of particle physics. Many lingering physics problems are linked to the Higgs boson, for instance, the stability of the vacuum, electroweak hierarchy problem and dark matter. These problems imply the existence of new physics beyond the SM and require a good understanding of the Higgs properties.
The effective coupling of the Higgs boson to a gluon pair is one of the most important parameters.
Many theories beyond the SM predict that the Higgs boson-gluon coupling may have deviation from the SM prediction by direct or indirect effects, for example, the stop in supersymmetry or the $T$ quark in little Higgs models can contribute to the coupling through the loop effects \cite{Blum:2012ii,Han:2013ic,Einhorn:1993hj,Kanemura:2004mg,He:2013tia,Moyotl:2016fdk,Baek:2017kxh,Hou:2017vvp,Kanemura:2017wtm,Passehr:2017ufr}.
Therefore, the precision measurement of the Higgs boson-gluon coupling will be a touchstone of the SM and may lead to a breakthrough for new physics.

Although the gluon fusion is the most important process of the Higgs boson production at the CERN Large Hadron Collider, the Higgs boson-gluon coupling is still difficult to be determined accurately due to the overwhelming large QCD radiation \cite{Peskin:2012we,Peskin:2013xra}.
The better candidates for the precision measurement of Higgs boson-gluon coupling can be electron positron colliders, which have the clean environment and the high luminosity.
The possible future electron positron colliders, which are usually called the Higgs factory at 250 GeV center-of-mass energy, include the Circular Electron-Positron Collider \cite{CEPCStudyGroup:2018rmc,CEPCStudyGroup:2018ghi,Mo:2015mza},
Future Circular Collider-electron-positron \cite{Gomez-Ceballos:2013zzn,Barletta:2014vea, Benedikt:2016vzy}
and International Linear Collider \cite{Behnke:2013xla,Baer:2013cma,Adolphsen:2013jya,Adolphsen:2013kya,Behnke:2013lya}.
At the Higgs factory, the measurement on most of the Higgs properties can reach percent level accuracy \cite{Peskin:2012we,Peskin:2013xra,Gao:2016jcm}.
For the Higgs boson-gluon effective coupling the $\kappa_g$ \cite{He:2013tia,CEPCStudyGroup:2018ghi} is always used to parameterize its deviation from the SM prediction, where $\kappa_g^{\rm SM}=1$.
With the conventional method (only using the kinematic cuts and $b$ tagging) \cite{YuBai} the uncertainty of the $\kappa_g$ will reach about $2.2\%$ for the channel of a $Z$ boson decaying to a lepton pair including the detector effect at the Circular Electron-Positron Collider.

The measurement accuracy of the Higgs boson-gluon coupling can be further improved through an effective identification of jet types.
In the last few decades, many different observables motivated by color charge, color connections, electrical charge, or spin have been proposed and achieved good performance \cite{Gallicchio:2012ez,Shelton:2013an,Larkoski:2017jix}.
For example, the jet energy profile is one of the useful jet substructure observables to distinguish quark and gluon jets by the energy distribution of jet constituents.
By using the jet energy profile, the uncertainty of the Higgs boson-gluon coupling can be further reduced to about $1.6\%$ for the channel of a $Z$ boson decaying to a lepton pair \cite{Li:2018qiy}.

However, an observable usually only describes a certain aspect of the jets or some special processes.
Although it is better to choose a set of complementary observables to extract more comprehensive characteristics to identify different types of jets or events, the applicable scope of different observables and the degree of association between them will also be difficult problems. Moreover, the deeper correlations between the jet or event constituents may be difficult to be extracted by the artificial observables.

Deep learning has been applied to solve many complicated problems in particle physics. In particular, deep neural networks have been employed to distinguish different types of jets, including Higgs boson tagging \cite{Chiappetta:1993zv}, boosted $W$ boson tagging \cite{Cogan:2014oua,deOliveira:2015xxd}, boosted top tagging \cite{Almeida:2015jua,Pearkes:2017hku}, single merged jet tagging \cite{Baldi:2016fql}, heavy-light quark discrimination \cite{Guest:2016iqz} and quark-gluon discrimination \cite{Lonnblad:1990bi,Lonnblad:1990qp,Peterson:1993nk,Komiske:2016rsd}.
They all get an exciting recognition capability and superior to the conventional method.
A convolutional neural network (CNN) is one of the most popular and powerful algorithms. Its powerful ability of image recognition makes it easy to extract more comprehensive and deeper features to analyze the jet substructure. It is very suitable for jet tagging and also for testing different shower and hadronization schemes by comparing different Monte Carlo (MC) generators.

In this paper, we propose to use the CNN for the precision measurement of Higgs boson-gluon effective coupling by distinguishing the background processes from the process of a $Z$ boson decaying to a lepton pair and a Higgs boson decaying to a gluon pair ($2\ell2g$) at lepton colliders.
The global information in an event is used for the training of the CNN instead of the jet information.
We will use events from different event generators for neural network training and testing to illuminate the difference between the different shower and hadronization schemes.

The content is organized as follows. In the next section, the CNN is briefly reviewed. In the third section, the MC events are generated by PYTHIA and HERWIG. The production of images and CNN architecture are introduced in the fourth section. In the fifth section, we show the results using the CNN. The conclusion is made in the last section.

\section{Convolutional Neural Networks}

A neural network is one of the most popular algorithms in machine learning. Generally, a neural network consists of an input layer, hidden layer, and output layer. A layer is dense if each of its units connects to all of the units in the previous layer. If a neural network consists of a dense layer completely, it will tune a large number of parameters and waste a lot of computing resources.
Actually, each neuron only needs to perceive the local image instead of the global image for image recognition, and then the global information can be obtained by integrating the local information at a higher level.
This motivates the design of the CNN \cite{3a592b7a61c24ba7a7a35bf0983c9b68}. In the last few years, based on the development of computer technology, the CNN has been a mainstay of many major breakthroughs in various fields.

In the image identification, the images in the CNN will pass a convolutional layer, pooling layer, and dense layer. The function of the convolutional layer is extracting features of the image. This can be implemented by the convolution of the filter and the image. A filter is a $n \times n$ grid of weights, where $n$ is the filter size. The convolution is that each weight in a filter multiplies the corresponding pixel intensity in a patch the same size as an image. Then, we sum the convolutional values, add a bias, and feed it to an activation function.
Activation functions introduce the nonlinear properties into neural networks, which enable the neural networks to learn the deeper information.
The most used activation function in CNNs is rectified linear units (ReLU), which is defined as $f(x)=\text{max}\{0,x\}$.
Each convolutional layer usually has many different filters to extract different features of an image.
For the multichannel images, there are different colors and convolutional filters in each channel. Each color or channel will be solved by a corresponding filter, like the single color image, and will be accumulated in the final step.

Then, a pooling layer, following the convolutional layer, is used to reduce the number of parameters.
The filter of the pooling layer is a $m \times m$ grid, where $m$ is the pooling size.
The max pooling and average pooling are the most common pooling functions. Max pooling takes the largest value while average pooling takes the average of all values in a filter region.
A dropout usually is added to avoid the overfitting. It refers to the randomly discarding of some neural network units at certain probability in each training \cite{JMLR:v15:srivastava14a}.
Finally, the dense layers are added to integrate the features in the feature maps extracted by the convolution layers and pooling layers to obtain the high-level meanings of the features and then use them for image recognition.

The error of the model can be quantified by the binary cross entropy loss function \cite{chollet2015keras}
\begin{align}
f_{\rm loss}=-\frac{1}{N}\sum_{i=1}^N[y_i \ln Y_i + (1-y_i)\ln(1-Y_i)],
\end{align}
where $N$ is the number of training events. The $y_i$ and $Y_i$ are the real value and the predicted value by the CNN of the $i$th event.
The training process is tuning the parameters in the model to minimize the loss function.

\section{Pre-Processing}

The main process of the Higgs boson production is $e^+e^- \rightarrow Z^* / \gamma^* \rightarrow Zh$ at the future $e^+e^-$ colliders. We choose the process of the $Z$ boson decaying to a lepton pair and the Higgs boson decaying to a gluon pair ($2\ell2g$) as the signal process since the $Z$ boson can be reconstructed very well by the lepton pair. The process of different $Z$ boson decay modes $Z\to e^{+}e^{-}$ and $Z\to \mu^{+}\mu^{-}$ are discussed first. Then the two lepton channels are combined as $Z\rightarrow \ell^{+}\ell^{-}$. The backgrounds are divided into two-fermion leptonic (final states are a lepton pair from the $Z$ or $\gamma^*$ intermediate states), two-fermion hadronic (final states are two quarks), four-fermion leptonic (final states are four leptons from the vector boson pair intermediate states), four-fermion semileptonic (final states are a pair of charged leptons and a pair of quarks from the vector boson pair intermediate states), four-fermion hadronic (final states are four quarks), and the Higgs boson production with the final states, which are different from the signal [mainly the Higgs boson and a $Z$ boson associated production process with the $Z$ boson decaying to a lepton pair and the Higgs boson decaying to a $b/c$ quark pair ($hbb/hcc$) or $W/Z$ boson pair ($hWW/hZZ$)] \cite{Yan:2016xyx,XinMo}.
Both the signal and background events are simulated at future $e^+e^-$ colliders \cite{CEPCStudyGroup:2018rmc,CEPCStudyGroup:2018ghi,Mo:2015mza,Gomez-Ceballos:2013zzn,Barletta:2014vea, Benedikt:2016vzy,Behnke:2013xla,Baer:2013cma,Adolphsen:2013jya,Adolphsen:2013kya,Behnke:2013lya}
for the center-of-mass energy 250 GeV and integrated luminosity 5 ab$^{-1}$.
The parton level MC events are generated by WHIZARD 1.95 \cite{Kilian:2007gr,Moretti:2001zz} and transferred to hadron level by PYTHIA 6 \cite{Sjostrand:2006za} and HERWIG 7 \cite{Bellm:2015jjp}, respectively. For clarity, we call them PYTHIA data and HERWIG data, respectively.

We select a pair of isolated leptons to reconstruct the $Z$ boson. The rest of the final state constituents are clustered into jets via FASTJET 3.3.0 \cite{Cacciari:2011ma} using the anti-$k_T$ algorithm with a large jet cone of $R=1.5$, and the energy of each jet is required to be more than 5 GeV.
To suppress the two-fermion leptonic and four-fermion leptonic backgrounds \cite{XinMo}, we add two cuts at first. One is the number of the stable charge particles in the final state $N_{\rm charge} \geq 10$, and another is the electromagnetic energy ratio in the final state $R_{\rm EM} < 0.99$. Then, the kinematic cuts, i.e., invariant mass, recoil mass, and other constraints of the lepton pair and jet pair, are used to ensure that the lepton pair and jet pair, respectively, come from the $Z$ boson and the Higgs boson to reject the two-fermion hadronic and four-fermion hadronic backgrounds. More details of the analysis can be found in Ref.\cite{Li:2018qiy}. The reference also shows that the $c$ tagging cannot decrease the $\kappa_g$ uncertainty effectively since its mistag rate for the gluon jet will exclude some gluon jets. Therefore, we only use the $b$ tagging in this paper.

The kinematic cuts and $b$ tagging can remove a large number of the distinct backgrounds, which will greatly improve the efficiency of the neural network.
The remaining backgrounds contain the $hbb$, $hcc$, $hWW$, $hZZ$ and four-fermion semileptonic. The jets in the backgrounds $hbb/hcc$ and four-fermion semileptonic are mainly heavy quark jets and light quark jets, respectively. But the jets in the backgrounds $hWW/hZZ$ are $W/Z$ jets and light quark jets since quite a few of the light quark jets are merged into the $W/Z$ jets with a large jet cone of $R=1.5$. It is the complex jet types in the backgrounds that make the signal identification be a challenge.

After all the cuts, the uncertainties of $\kappa_g$ should be evaluated.
The evaluation of systematic uncertainties requires a detailed detector study and is unknown yet for the Higgs factory.
But the statistical uncertainty of $\kappa_g$ around the SM prediction can be explicitly expressed as
\begin{align}
\delta \kappa_g=\frac{\sqrt{N}}{2N_g},
\end{align}
where $N_g$ and $N$ are the numbers of the Higgs boson decaying to gluon pair events and total events, respectively.

\begin{table}[!htb]
\centering
\begin{tabular}{cccc}
\toprule
 & $\quad Z\rightarrow e^{+}e^{-} \quad$ & $\quad Z\rightarrow \mu^{+}\mu^{-} \quad$ & $\quad Z\rightarrow \ell^{+}\ell^{-}$ \\
\hline
PYTHIA  & 2.93\% & 2.53\% & 1.94\% \\
HERWIG  & 2.67\% & 2.47\% & 1.82\% \\
\botrule
\end{tabular}
\caption{The uncertainties of $\kappa_g$ in the different $Z$ boson decay modes with the conventional method using PYTHIA data and HERWIG data. }
\label{kappag}
\end{table}

In Table \ref{kappag}, the second and the third lines are the uncertainties of $\kappa_g$ with the conventional method using PYTHIA data and HERWIG data, respectively. The difference between the results using PYTHIA data and HERWIG data may come from the different shower and hadronization schemes. The $k_{T}$-ordered and the angular-ordered schemes are used for shower effect, and the Lund string and the cluster models are used for the hadronization effect in PYTHIA 6 and HERWIG 7, respectively.

\section{Architecture of CNN}

For the training of the CNN, we use the combined lepton channel $Z\rightarrow \ell^{+}\ell^{-}$.
The entire spherical surface, where the azimuthal angle $\phi \in [-\pi,\pi]$ and the polar angle $\theta \in [0,\pi]$, is treated as a two dimensional plane image. Each image is designed to have a 66-pixel length in the $\phi$ direction and a 34-pixel length in the $\theta$ direction. The energy of all the final state stable particles is discretized into pixels as our pixel intensity at lepton colliders.
The images of the signal process $2\ell2g$ are given the sign one and the other images as the background process are given the sign zero.
All the images are divided into the training, validation and test sets in proportion to 8:1:1.

\begin{figure}[!htb]
\centering
\includegraphics[scale=0.25]{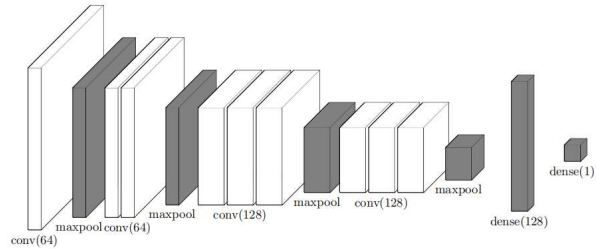}
\caption{\label{FIG:cnn} The architecture of our CNN.}
\end{figure}

The neural network is implemented by using Keras \cite{chollet2015keras} with TensorFlow backend. Our CNN architecture is inspired by the VGGNet \cite{Simonyan14c} architectures and consisted of four iterations of convolutional layers and maxpooling layers shown in Fig.\ref{FIG:cnn}. Then the feature map is flattened and fed to a dense layer with 128 units. Finally, a dense layer with one unit and a sigmoid activation is added to classify the signal and background processes.
Each convolutional layer consists of 64 or 128 filters with filter size $3 \times 3$ and a ReLU activation. The uniform distribution is used to initialize the filters. The stride length of the convolution is 1. The first convolutional layer is set without padding to weaken the influence of the edge information of the image at the beginning while the others are set with padding to keep all the information of the feature map.
Each maxpooling layer performs a $2 \times 2$ down-sampling with a stride length of 2.
A dropout layer follows each maxpooling layer and the dense layer to avoid overfitting. All the dropout rates of dropout layers are 0.5 except that the first one is 0.25.

The binary cross entropy is used as the loss function.
The optimization of training uses the Adam algorithm \cite{Kingma:2014vow} and the learning rate is 0.0005.
The training is set with batch size 128 and 100 epochs and an early stopping patience of 5.
Thus, the training will stop early if the value of the validation loss does not go down 5 times
\footnote{Example code is provided on https://github.com/zhaoli-IHEP/Higgs-ML.}.

The receiver operator characteristic (ROC) curve is usually used to quantify the performance of neural networks. A ROC curve is generated by plotting the true positive rate against the false positive rate. The area under the curve (AUC) is defined to compare the overall performance of the neural networks.
In this paper, the true positive rate is the signal process ($2\ell2g$) acceptance efficiency $R_g$ and the false positive rate is the mistag efficiency $R_B$ of the background processes.

\begin{figure}[!htb]
\centering
\includegraphics[scale=0.35]{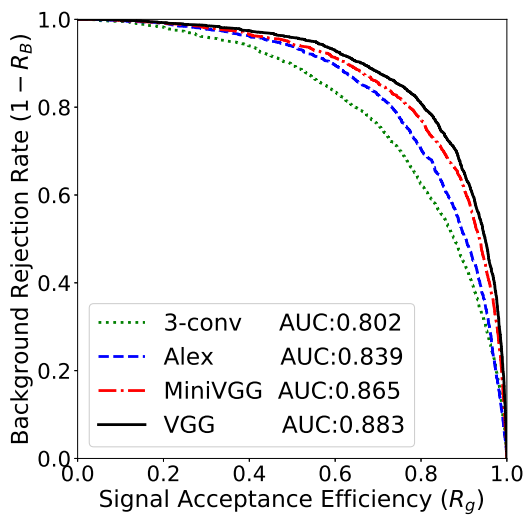}
\caption{\label{FIG:net}
The background rejection rate $1-R_B$ as a function of the signal acceptance efficiency $R_g$ for the CNN with different architectures. }
\end{figure}

Then we test the performance of our neural network and compare it to several different neural networks.
Fig.\ref{FIG:net} shows the background rejection rate $1-R_B$ as a function of the signal acceptance efficiency $R_g$ for the CNN with different architectures.
The lines marked as "3-conv", "Alex" and "MiniVGG" represent the performance of the CNN architectures in the Refs.\cite{Komiske:2016rsd,NIPS2012_4824,Guo:2018hbv}, respectively. The green dotted line is the result using the neural network, which contains three iterations of a convolutional layer and a maxpooling layer. The blue dash line is the result using the famous AlexNet, which uses a stack of convolutional layers to increase the nonlinearity of the neural network and bigger filter size to increase the receptive field. So, the performance of the AlexNet has a significant improvement compared to that of the 3-conv. The red dash-dotted line is the result using the neural network, which is inspired by the MiniVGGNet architecture but with a bigger filter size in the first two convolutional layers. More iterations of the convolution layer stack further enhance the nonlinearity of the neural network and lead to improved performance.
According to the advantages of the VGGNet, our neural network uses a stack of convolutional layers with $3\times3$ filter size instead of a single convolutional layer with a big filter size, which can increase the nonlinearity of the neural network and reduce the number of parameters.
The black solid line is the result using our neural network architecture, which is better than other three neural network structures for the identification of our signal and background processes.

\section{Results}

In this section, we will present the improvement on the $\kappa_g$ uncertainty archived by using the CNN.

\begin{figure}[!htb]
\centering
\includegraphics[scale=0.35]{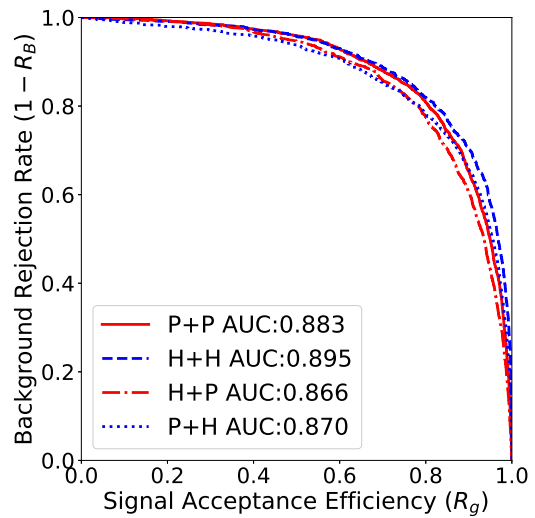}
\caption{\label{FIG:roc}
The background rejection rate $1-R_B$ as a function of the signal acceptance efficiency $R_g$ for our CNN. The symbol "P(H)+P(H)" means training with the PYTHIA (HERWIG) data and testing with the PYTHIA (HERWIG) data. }
\end{figure}

Fig.\ref{FIG:roc} shows the background rejection rate $1-R_B$ as a function of the signal acceptance efficiency $R_g$ for our CNN. The area under these curves are the AUC values of the different cases. Both training and testing have been applied to the PYTHIA and HERWIG data. For convenience, The symbol "P(H)+P(H)" is used to represent training with the PYTHIA (HERWIG) data and testing with the PYTHIA (HERWIG) data.
It can be found that at around $R_g=80\%$ the background rejection rate can reach about $80\%$, meanwhile, the signal acceptance efficiency could still be acceptable.
Furthermore, it can be seen that the AUC value of the "H+H" is slightly better than that of the "P+P". More specifically, the curves of the P+P and H+H are very similar at the low signal acceptance efficiency region $R_g<70\%$, but the curve of the H+H is higher than that of the P+P at the high signal acceptance efficiency region $R_g>70\%$.
In general, the performance of the P+P and H+H are similar, which indicates that the similar performance of the shower and hadronization schemes in PYTHIA and HERWIG.

The "H+P" and "P+H" are training and testing with different data as a cross-check to illustrate the universality of the CNN model.
It makes sense to compare the performance of the CNN models, which are trained with the different data but tested with the same data.
The CNN models are universal if their performance are similar.
By comparing the "P+P(H)" to the "H+P(H)" in Fig.\ref{FIG:roc}, the performance of the CNN model tested with different data is just slightly worse than that tested with same data in all the signal acceptance efficiency region. It means that our CNN models do not have too much overfitting since they are not overly dependent on the certain data.

The different ratios of the remaining signal and backgrounds can be obtained on the ROC curve in Fig.\ref{FIG:roc}. The uncertainty of $\kappa_g$ after using the CNN at each point $(R_g, R_B)$ can be expressed as
\begin{align}
\delta \kappa_g^{\rm CNN}(R_g, R_B)=\frac{\sqrt{N_g R_g+N_B R_B}}{2N_g R_g}.
\end{align}

\begin{figure}[!htb]
\centering
\includegraphics[scale=0.37]{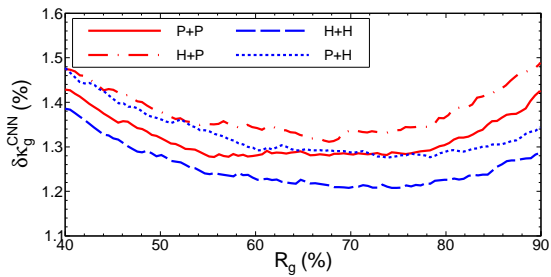}
\caption{\label{FIG:norotation}
The uncertainty of $\kappa_g$ after the CNN as a function of the signal acceptance efficiency $R_g$. Both training and testing use the PYTHIA and HERWIG data.}
\end{figure}

Fig.\ref{FIG:norotation} presents the uncertainty of $\kappa_g$ after CNN as a function of the signal acceptance efficiency $R_g$ using the PYTHIA and HERWIG data.
At the optimal point $R_g=70\%$, $\delta\kappa_g^{\rm CNN}$ can reach about $1.28\%$ by using the P+P and $1.22\%$ by using the H+H.
Compared to Table \ref{kappag}, it shows that $\delta\kappa_g^{\rm CNN}$ can be further reduced by $34\%$ for the P+P and $33\%$ for the H+H.
The results using the H+H is about $5\%$ smaller than that using the P+P. The small difference of the results may come from the different shower and hadronization schemes in PYTHIA and HERWIG.
The results of the cross check are slightly worse than that of the training and testing with the same data.
Comparing the P+P to the H+P, the uncertainties of $\kappa_g$ using the H+P is slightly worse than that using the P+P.
But the difference of the P+P and the H+P is less than $0.1\%$, which far exceeds the measurement accuracy of the future electron positron colliders. The H+H and the P+H are in the same situation. The similar results mean that our CNN models do not have too much overfitting and the results are reliable.

\begin{figure}[!htb]
\centering
\includegraphics[scale=0.37]{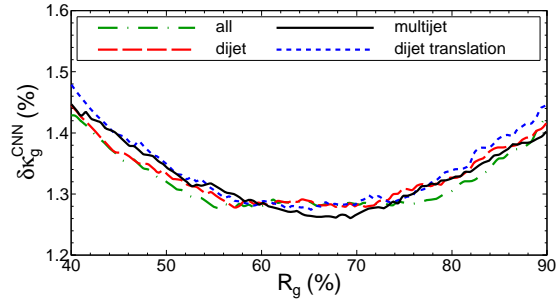}
\caption{\label{FIG:multijet}
The uncertainty of $\kappa_g$ after the CNN as a function of the signal acceptance efficiency $R_g$ using the different images, which are constructed with the information of all the final state stable particles, all the jets, or the first two jets sorted by their energy. }
\end{figure}

In the previous part, one image is constructed with the information of all the final state stable particles in an event.
To gain insight into the improvement by the CNN and find the most important features of the signal and background, different images are constructed with different final state constituents. The following analysis only uses the PYTHIA data.
Fig.\ref{FIG:multijet} shows the uncertainty of $\kappa_g$ after the CNN as a function of the signal acceptance efficiency $R_g$ using the different images.
The line marked as "all" is the result using the images constructed with the information of all the final state stable particles, and the line marked as "multijet" is the result using the images constructed with the information of all the jets clustered by anti-$k_T$ algorithm in an event.
The "multijet" result is slightly better than the "all" result in the region $R_g\in[60\%, 70\%]$. However, the difference of the "all" and "multijet" results is less than $0.2\%$ at the optimal points and can be ignored. This indicates that the information of jets makes a major contribution to the identification of the signal and background processes. The reason is that most of the information except the jets in an event is the lepton pair, which are very similar in the signal and background processes after using the kinematic cuts.
The line marked as "dijet" is the result using the images only constructed with the information of the leading and subleading jets.
The "all" and "dijet" results are very similar, which shows that the leading and subleading jets nearly contribute all the features for the CNN.
The "multijet" and "dijet" results are also very similar since most of the events only have two jets with a large jet cone of $R=1.5$.
If the images are constructed only with the leading and subleading jets, the center of the two jets can be chosen as the image center. Then the constituents of the two jets are discretized into pixels to obtain the "dijet translation" images. By this operation, the jets will not be split into two parts at the margins of the image. It can be seen that the "dijet translation" and the "dijet" results are also very similar, which indicates that the symmetry property in the $\phi$ direction has been recognized by the CNN.

\begin{figure}[!htb]
\centering
\includegraphics[scale=0.37]{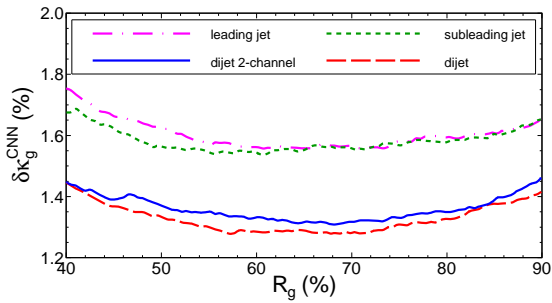}
\caption{\label{FIG:singlejet}
The uncertainty of $\kappa_g$ after the CNN varies with the signal acceptance efficiency $R_g$ using different single-jet images, which are constructed only with the leading jet or subleading jet. }
\end{figure}

After showing that the information of the leading and subleading jets makes a major contribution to the identification of the signal and background processes, we further analyze the contribution of each jet.
Fig.\ref{FIG:singlejet} shows the uncertainty of $\kappa_g$ after the CNN as a function of the signal acceptance efficiency $R_g$ using the different single-jet images. Each single-jet image has the size $2R \times 2R$ with the jet cone $R=1.5$ and is designed to have $34 \times 34$ pixels. The jet axis is chosen at the image center so that there is a complete jet on the single-jet image.
The lines marked as "leading jet" and "subleading jet" represent the results using the leading jet images and the subleading jet images, respectively.
We can see that the leading and subleading jets are equally important for the identification.
Then the leading and subleading jet images as two different channels are combined as the "dijet 2-channel" by analogy with the recognition of color images, with red, green and blue intensities treated as separate input layers. Compared to the "dijet", which puts the leading and subleading jets in one image, the "dijet 2-channel" removes the relative location information of the two jets.
It can be seen that the "dijet 2-channel" result is just slightly worse than the "dijet" result, so the relative location information of the jets is not important for this discrimination.
From the above analysis, we can conclude that the leading and subleading jets make a major contribution to the identification of the signal and background processes.

In the third section, the analysis shows that the jets in the signal process are mainly gluon jets and the jets in the background processes can be mainly divided into quark jets and $W/Z$ jets.
The three types of jets have different energy distributions of their constituents. Each jet image using energy as pixel intensity records the energy distribution of the jet constituents. This information can be extracted from the jet images by the CNN to identify the signal and background processes. Therefore, the energy distributions of the leading and subleading jets constituents make a major contribution to the identification of the signal and background processes.

\begin{figure}[!htb]
\centering
\includegraphics[scale=0.37]{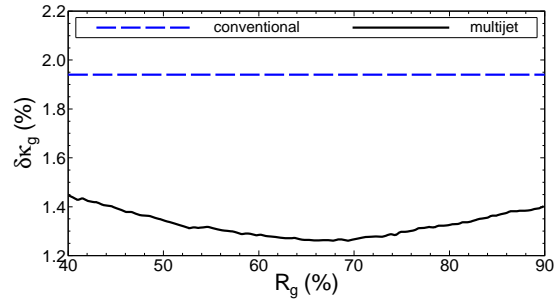}
\caption{\label{FIG:best}
The best result using the CNN is compared to the result using the conventional method for the PYTHIA data. }
\end{figure}

Fig.\ref{FIG:best} shows the best result using the CNN (the line marked as "multijet") and the result using the conventional method (the line marked as "conventional") for the PYTHIA data.
Comparing to the result using the conventional method, the CNN has a significant improvement in a wide signal acceptance efficiency region.
At the optimal point $R_g=70\%$, the uncertainty of $\kappa_g$ can be decreased from $1.94\%$ to about $1.26\%$ by using the CNN and reduced by about $35\%$ compared to that using the conventional method for the PYTHIA data. Moreover, the result using the HERWIG data is similar to that using the PYTHIA data.

\section{Conclusions}

In this paper, the CNN is used to improve the precision measurement of the Higgs boson-gluon effective coupling at lepton colliders.
By using the CNN the uncertainty of $\kappa_g$ can be decreased from $1.94\%$ to about $1.28\%$ using the PYTHIA data
and from $1.82\%$ to about $1.22\%$ using the HERWIG data in the channel of a $Z$ boson decaying to a lepton pair
in the MC simulation for the center-of-mass energy 250 GeV and integrated luminosity 5 ab$^{-1}$.
The difference between the expected $\kappa_g$ uncertainties using the PYTHIA and the HERWIG data is less than $0.1\%$.
Moreover, the performance of the CNN using different final state constituents is proof
that the energy distributions of the leading and subleading jets constituents play a major role on the identification
and the optimal uncertainty of $\kappa_g$ using the CNN is reduced by about $35\%$ compared to that using the conventional method.

\vspace{1ex}
This work was supported by the National Natural Science Foundation of China under Grant No. 11675185.
Y.W. is supported by the China Postdoctoral Science Foundation under Grant No. 2016M601134.
The authors want to thank Yu Bai, Gang Li, Qiang Li and Manqi Ruan for helpful discussions.

\bibliography{CNN}

\begin{thebibliography}{54}
\expandafter\ifx\csname natexlab\endcsname\relax\def\natexlab#1{#1}\fi
\expandafter\ifx\csname bibnamefont\endcsname\relax
  \def\bibnamefont#1{#1}\fi
\expandafter\ifx\csname bibfnamefont\endcsname\relax
  \def\bibfnamefont#1{#1}\fi
\expandafter\ifx\csname citenamefont\endcsname\relax
  \def\citenamefont#1{#1}\fi
\expandafter\ifx\csname url\endcsname\relax
  \def\url#1{\texttt{#1}}\fi
\expandafter\ifx\csname urlprefix\endcsname\relax\def\urlprefix{URL }\fi
\providecommand{\bibinfo}[2]{#2}
\providecommand{\eprint}[2][]{\url{#2}}

\bibitem[{\citenamefont{Blum et~al.}(2013)\citenamefont{Blum, D'Agnolo, and
  Fan}}]{Blum:2012ii}
\bibinfo{author}{\bibfnamefont{K.}~\bibnamefont{Blum}},
  \bibinfo{author}{\bibfnamefont{R.~T.} \bibnamefont{D'Agnolo}},
  \bibnamefont{and} \bibinfo{author}{\bibfnamefont{J.}~\bibnamefont{Fan}},
  \bibinfo{journal}{JHEP} \textbf{\bibinfo{volume}{01}}, \bibinfo{pages}{057}
  (\bibinfo{year}{2013}), \eprint{1206.5303}.

\bibitem[{\citenamefont{Han et~al.}(2013)\citenamefont{Han, Wang, Yang, and
  Zhu}}]{Han:2013ic}
\bibinfo{author}{\bibfnamefont{X.-F.} \bibnamefont{Han}},
  \bibinfo{author}{\bibfnamefont{L.}~\bibnamefont{Wang}},
  \bibinfo{author}{\bibfnamefont{J.~M.} \bibnamefont{Yang}}, \bibnamefont{and}
  \bibinfo{author}{\bibfnamefont{J.}~\bibnamefont{Zhu}},
  \bibinfo{journal}{Phys. Rev.} \textbf{\bibinfo{volume}{D87}},
  \bibinfo{pages}{055004} (\bibinfo{year}{2013}), \eprint{1301.0090}.

\bibitem[{\citenamefont{Einhorn}(1993)}]{Einhorn:1993hj}
\bibinfo{author}{\bibfnamefont{M.~B.} \bibnamefont{Einhorn}}, in
  \emph{\bibinfo{booktitle}{{Conference on Unified Symmetry in the Small and in
  the Large Coral Gables, Florida, January 25-27, 1993}}}
  (\bibinfo{year}{1993}), pp. \bibinfo{pages}{407--420},
  \eprint{hep-ph/9303323}.

\bibitem[{\citenamefont{Kanemura et~al.}(2004)\citenamefont{Kanemura, Okada,
  Senaha, and Yuan}}]{Kanemura:2004mg}
\bibinfo{author}{\bibfnamefont{S.}~\bibnamefont{Kanemura}},
  \bibinfo{author}{\bibfnamefont{Y.}~\bibnamefont{Okada}},
  \bibinfo{author}{\bibfnamefont{E.}~\bibnamefont{Senaha}}, \bibnamefont{and}
  \bibinfo{author}{\bibfnamefont{C.~P.} \bibnamefont{Yuan}},
  \bibinfo{journal}{Phys. Rev.} \textbf{\bibinfo{volume}{D70}},
  \bibinfo{pages}{115002} (\bibinfo{year}{2004}).

\bibitem[{\citenamefont{He et~al.}(2013)\citenamefont{He, Tang, and
  Valencia}}]{He:2013tia}
\bibinfo{author}{\bibfnamefont{X.-G.} \bibnamefont{He}},
  \bibinfo{author}{\bibfnamefont{Y.}~\bibnamefont{Tang}}, \bibnamefont{and}
  \bibinfo{author}{\bibfnamefont{G.}~\bibnamefont{Valencia}},
  \bibinfo{journal}{Phys. Rev.} \textbf{\bibinfo{volume}{D88}},
  \bibinfo{pages}{033005} (\bibinfo{year}{2013}), \eprint{1305.5420}.

\bibitem[{\citenamefont{Moyotl et~al.}(2016)\citenamefont{Moyotl, Chamorro,
  Castilla-Valdez, and Pérez}}]{Moyotl:2016fdk}
\bibinfo{author}{\bibfnamefont{A.}~\bibnamefont{Moyotl}},
  \bibinfo{author}{\bibfnamefont{S.}~\bibnamefont{Chamorro}},
  \bibinfo{author}{\bibfnamefont{H.}~\bibnamefont{Castilla-Valdez}},
  \bibnamefont{and} \bibinfo{author}{\bibfnamefont{M.~A.} \bibnamefont{Pérez}}
  (\bibinfo{year}{2016}), \eprint{1610.06299}.

\bibitem[{\citenamefont{Baek and Yuan}(2017)}]{Baek:2017kxh}
\bibinfo{author}{\bibfnamefont{S.}~\bibnamefont{Baek}} \bibnamefont{and}
  \bibinfo{author}{\bibfnamefont{X.-B.} \bibnamefont{Yuan}},
  \bibinfo{journal}{Phys. Lett.} \textbf{\bibinfo{volume}{B774}},
  \bibinfo{pages}{662} (\bibinfo{year}{2017}).

\bibitem[{\citenamefont{Hou and Kikuchi}(2017)}]{Hou:2017vvp}
\bibinfo{author}{\bibfnamefont{W.-S.} \bibnamefont{Hou}} \bibnamefont{and}
  \bibinfo{author}{\bibfnamefont{M.}~\bibnamefont{Kikuchi}},
  \bibinfo{journal}{Phys. Rev.} \textbf{\bibinfo{volume}{D96}},
  \bibinfo{pages}{015033} (\bibinfo{year}{2017}).

\bibitem[{\citenamefont{Kanemura et~al.}(2017)\citenamefont{Kanemura, Kikuchi,
  Sakurai, and Yagyu}}]{Kanemura:2017wtm}
\bibinfo{author}{\bibfnamefont{S.}~\bibnamefont{Kanemura}},
  \bibinfo{author}{\bibfnamefont{M.}~\bibnamefont{Kikuchi}},
  \bibinfo{author}{\bibfnamefont{K.}~\bibnamefont{Sakurai}}, \bibnamefont{and}
  \bibinfo{author}{\bibfnamefont{K.}~\bibnamefont{Yagyu}},
  \bibinfo{journal}{Phys. Rev.} \textbf{\bibinfo{volume}{D96}},
  \bibinfo{pages}{035014} (\bibinfo{year}{2017}).

\bibitem[{\citenamefont{Paßehr and Weiglein}(2017)}]{Passehr:2017ufr}
\bibinfo{author}{\bibfnamefont{S.}~\bibnamefont{Paßehr}} \bibnamefont{and}
  \bibinfo{author}{\bibfnamefont{G.}~\bibnamefont{Weiglein}}
  (\bibinfo{year}{2017}), \eprint{1705.07909}.

\bibitem[{\citenamefont{Peskin}(2012)}]{Peskin:2012we}
\bibinfo{author}{\bibfnamefont{M.~E.} \bibnamefont{Peskin}}
  (\bibinfo{year}{2012}), \eprint{1207.2516}.

\bibitem[{\citenamefont{Peskin}(2013)}]{Peskin:2013xra}
\bibinfo{author}{\bibfnamefont{M.~E.} \bibnamefont{Peskin}}, in
  \emph{\bibinfo{booktitle}{{Proceedings, 2013 Community Summer Study on the
  Future of U.S. Particle Physics: Snowmass on the Mississippi (CSS2013):
  Minneapolis, MN, USA, July 29-August 6, 2013}}} (\bibinfo{year}{2013}),
  \eprint{1312.4974}.

\bibitem[{\citenamefont{{CEPC Study
  Group}}(2018{\natexlab{a}})}]{CEPCStudyGroup:2018rmc}
\bibinfo{author}{\bibnamefont{{CEPC Study Group}}}
  (\bibinfo{year}{2018}{\natexlab{a}}), \eprint{1809.00285}.

\bibitem[{\citenamefont{{CEPC Study
  Group}}(2018{\natexlab{b}})}]{CEPCStudyGroup:2018ghi}
\bibinfo{author}{\bibnamefont{{CEPC Study Group}}}
  (\bibinfo{year}{2018}{\natexlab{b}}), \eprint{1811.10545}.

\bibitem[{\citenamefont{Mo et~al.}(2016)\citenamefont{Mo, Li, Ruan, and
  Lou}}]{Mo:2015mza}
\bibinfo{author}{\bibfnamefont{X.}~\bibnamefont{Mo}},
  \bibinfo{author}{\bibfnamefont{G.}~\bibnamefont{Li}},
  \bibinfo{author}{\bibfnamefont{M.-Q.} \bibnamefont{Ruan}}, \bibnamefont{and}
  \bibinfo{author}{\bibfnamefont{X.-C.} \bibnamefont{Lou}},
  \bibinfo{journal}{Chin. Phys.} \textbf{\bibinfo{volume}{C40}},
  \bibinfo{pages}{033001} (\bibinfo{year}{2016}).

\bibitem[{\citenamefont{Bicer et~al.}(2014)}]{Gomez-Ceballos:2013zzn}
\bibinfo{author}{\bibfnamefont{M.}~\bibnamefont{Bicer}} \bibnamefont{et~al.}
  (\bibinfo{collaboration}{TLEP Design Study Working Group}),
  \bibinfo{journal}{JHEP} \textbf{\bibinfo{volume}{01}}, \bibinfo{pages}{164}
  (\bibinfo{year}{2014}).

\bibitem[{\citenamefont{Barletta et~al.}(2014)\citenamefont{Barletta,
  Battaglia, Klute, Mangano, Prestemon, Rossi, and Skands}}]{Barletta:2014vea}
\bibinfo{author}{\bibfnamefont{W.}~\bibnamefont{Barletta}},
  \bibinfo{author}{\bibfnamefont{M.}~\bibnamefont{Battaglia}},
  \bibinfo{author}{\bibfnamefont{M.}~\bibnamefont{Klute}},
  \bibinfo{author}{\bibfnamefont{M.}~\bibnamefont{Mangano}},
  \bibinfo{author}{\bibfnamefont{S.}~\bibnamefont{Prestemon}},
  \bibinfo{author}{\bibfnamefont{L.}~\bibnamefont{Rossi}}, \bibnamefont{and}
  \bibinfo{author}{\bibfnamefont{P.}~\bibnamefont{Skands}},
  \bibinfo{journal}{Nucl. Instrum. Meth.} \textbf{\bibinfo{volume}{A764}},
  \bibinfo{pages}{352} (\bibinfo{year}{2014}).

\bibitem[{\citenamefont{Benedikt and Zimmermann}(2016)}]{Benedikt:2016vzy}
\bibinfo{author}{\bibfnamefont{M.}~\bibnamefont{Benedikt}} \bibnamefont{and}
  \bibinfo{author}{\bibfnamefont{F.}~\bibnamefont{Zimmermann}},
  \bibinfo{journal}{PoS} \textbf{\bibinfo{volume}{LeptonPhoton2015}},
  \bibinfo{pages}{052} (\bibinfo{year}{2016}).

\bibitem[{\citenamefont{Behnke et~al.}(2013)\citenamefont{Behnke, Brau, Foster,
  Fuster, Harrison, Paterson, Peskin, Stanitzki, Walker, and
  Yamamoto}}]{Behnke:2013xla}
\bibinfo{author}{\bibfnamefont{T.}~\bibnamefont{Behnke}},
  \bibinfo{author}{\bibfnamefont{J.~E.} \bibnamefont{Brau}},
  \bibinfo{author}{\bibfnamefont{B.}~\bibnamefont{Foster}},
  \bibinfo{author}{\bibfnamefont{J.}~\bibnamefont{Fuster}},
  \bibinfo{author}{\bibfnamefont{M.}~\bibnamefont{Harrison}},
  \bibinfo{author}{\bibfnamefont{J.~M.} \bibnamefont{Paterson}},
  \bibinfo{author}{\bibfnamefont{M.}~\bibnamefont{Peskin}},
  \bibinfo{author}{\bibfnamefont{M.}~\bibnamefont{Stanitzki}},
  \bibinfo{author}{\bibfnamefont{N.}~\bibnamefont{Walker}}, \bibnamefont{and}
  \bibinfo{author}{\bibfnamefont{H.}~\bibnamefont{Yamamoto}}
  (\bibinfo{year}{2013}), \eprint{1306.6327}.

\bibitem[{\citenamefont{Baer et~al.}(2013)\citenamefont{Baer, Barklow, Fujii,
  Gao, Hoang, Kanemura, List, Logan, Nomerotski, Perelstein
  et~al.}}]{Baer:2013cma}
\bibinfo{author}{\bibfnamefont{H.}~\bibnamefont{Baer}},
  \bibinfo{author}{\bibfnamefont{T.}~\bibnamefont{Barklow}},
  \bibinfo{author}{\bibfnamefont{K.}~\bibnamefont{Fujii}},
  \bibinfo{author}{\bibfnamefont{Y.}~\bibnamefont{Gao}},
  \bibinfo{author}{\bibfnamefont{A.}~\bibnamefont{Hoang}},
  \bibinfo{author}{\bibfnamefont{S.}~\bibnamefont{Kanemura}},
  \bibinfo{author}{\bibfnamefont{J.}~\bibnamefont{List}},
  \bibinfo{author}{\bibfnamefont{H.~E.} \bibnamefont{Logan}},
  \bibinfo{author}{\bibfnamefont{A.}~\bibnamefont{Nomerotski}},
  \bibinfo{author}{\bibfnamefont{M.}~\bibnamefont{Perelstein}},
  \bibnamefont{et~al.} (\bibinfo{year}{2013}), \eprint{1306.6352}.

\bibitem[{\citenamefont{Adolphsen
  et~al.}(2013{\natexlab{a}})\citenamefont{Adolphsen, Barone, Barish, Buesser,
  Burrows, Carwardine, Clark, Mainaud~Durand, Dugan, Elsen
  et~al.}}]{Adolphsen:2013jya}
\bibinfo{author}{\bibfnamefont{C.}~\bibnamefont{Adolphsen}},
  \bibinfo{author}{\bibfnamefont{M.}~\bibnamefont{Barone}},
  \bibinfo{author}{\bibfnamefont{B.}~\bibnamefont{Barish}},
  \bibinfo{author}{\bibfnamefont{K.}~\bibnamefont{Buesser}},
  \bibinfo{author}{\bibfnamefont{P.}~\bibnamefont{Burrows}},
  \bibinfo{author}{\bibfnamefont{J.}~\bibnamefont{Carwardine}},
  \bibinfo{author}{\bibfnamefont{J.}~\bibnamefont{Clark}},
  \bibinfo{author}{\bibfnamefont{H.}~\bibnamefont{Mainaud~Durand}},
  \bibinfo{author}{\bibfnamefont{G.}~\bibnamefont{Dugan}},
  \bibinfo{author}{\bibfnamefont{E.}~\bibnamefont{Elsen}}, \bibnamefont{et~al.}
  (\bibinfo{year}{2013}{\natexlab{a}}), \eprint{1306.6353}.

\bibitem[{\citenamefont{Adolphsen
  et~al.}(2013{\natexlab{b}})\citenamefont{Adolphsen, Barone, Barish, Buesser,
  Burrows, Carwardine, Clark, Mainaud~Durand, Dugan, Elsen
  et~al.}}]{Adolphsen:2013kya}
\bibinfo{author}{\bibfnamefont{C.}~\bibnamefont{Adolphsen}},
  \bibinfo{author}{\bibfnamefont{M.}~\bibnamefont{Barone}},
  \bibinfo{author}{\bibfnamefont{B.}~\bibnamefont{Barish}},
  \bibinfo{author}{\bibfnamefont{K.}~\bibnamefont{Buesser}},
  \bibinfo{author}{\bibfnamefont{P.}~\bibnamefont{Burrows}},
  \bibinfo{author}{\bibfnamefont{J.}~\bibnamefont{Carwardine}},
  \bibinfo{author}{\bibfnamefont{J.}~\bibnamefont{Clark}},
  \bibinfo{author}{\bibfnamefont{H.}~\bibnamefont{Mainaud~Durand}},
  \bibinfo{author}{\bibfnamefont{G.}~\bibnamefont{Dugan}},
  \bibinfo{author}{\bibfnamefont{E.}~\bibnamefont{Elsen}}, \bibnamefont{et~al.}
  (\bibinfo{year}{2013}{\natexlab{b}}), \eprint{1306.6328}.

\bibitem[{\citenamefont{Abramowicz et~al.}(2013)}]{Behnke:2013lya}
\bibinfo{author}{\bibfnamefont{H.}~\bibnamefont{Abramowicz}}
  \bibnamefont{et~al.} (\bibinfo{year}{2013}), \eprint{1306.6329}.

\bibitem[{\citenamefont{Gao}(2018)}]{Gao:2016jcm}
\bibinfo{author}{\bibfnamefont{J.}~\bibnamefont{Gao}}, \bibinfo{journal}{JHEP}
  \textbf{\bibinfo{volume}{01}}, \bibinfo{pages}{038} (\bibinfo{year}{2018}).

\bibitem[{\citenamefont{Bai}(2017)}]{YuBai}
\bibinfo{author}{\bibfnamefont{Y.}~\bibnamefont{Bai}}
  (\bibinfo{collaboration}{CEPC Working Group}),
  \emph{\bibinfo{title}{{Measurements of the decay branching fraction of $H\to
  b\bar b/c\bar c/gg$ at CEPC (CEPC Note in preparation)}}}
  (\bibinfo{year}{2017}).

\bibitem[{\citenamefont{Gallicchio and Schwartz}(2013)}]{Gallicchio:2012ez}
\bibinfo{author}{\bibfnamefont{J.}~\bibnamefont{Gallicchio}} \bibnamefont{and}
  \bibinfo{author}{\bibfnamefont{M.~D.} \bibnamefont{Schwartz}},
  \bibinfo{journal}{JHEP} \textbf{\bibinfo{volume}{04}}, \bibinfo{pages}{090}
  (\bibinfo{year}{2013}), \eprint{1211.7038}.

\bibitem[{\citenamefont{Shelton}(2013)}]{Shelton:2013an}
\bibinfo{author}{\bibfnamefont{J.}~\bibnamefont{Shelton}}, in
  \emph{\bibinfo{booktitle}{{Proceedings, Theoretical Advanced Study Institute
  in Elementary Particle Physics: Searching for New Physics at Small and Large
  Scales (TASI 2012): Boulder, Colorado, June 4-29, 2012}}}
  (\bibinfo{year}{2013}), pp. \bibinfo{pages}{303--340}, \eprint{1302.0260}.

\bibitem[{\citenamefont{Larkoski et~al.}(2017)\citenamefont{Larkoski, Moult,
  and Nachman}}]{Larkoski:2017jix}
\bibinfo{author}{\bibfnamefont{A.~J.} \bibnamefont{Larkoski}},
  \bibinfo{author}{\bibfnamefont{I.}~\bibnamefont{Moult}}, \bibnamefont{and}
  \bibinfo{author}{\bibfnamefont{B.}~\bibnamefont{Nachman}}
  (\bibinfo{year}{2017}), \eprint{1709.04464}.

\bibitem[{\citenamefont{Li et~al.}(2018)\citenamefont{Li, Li, Liu, Wang, and
  Zhao}}]{Li:2018qiy}
\bibinfo{author}{\bibfnamefont{G.}~\bibnamefont{Li}},
  \bibinfo{author}{\bibfnamefont{Z.}~\bibnamefont{Li}},
  \bibinfo{author}{\bibfnamefont{Y.}~\bibnamefont{Liu}},
  \bibinfo{author}{\bibfnamefont{Y.}~\bibnamefont{Wang}}, \bibnamefont{and}
  \bibinfo{author}{\bibfnamefont{X.}~\bibnamefont{Zhao}},
  \bibinfo{journal}{Phys. Rev.} \textbf{\bibinfo{volume}{D98}},
  \bibinfo{pages}{076010} (\bibinfo{year}{2018}), \eprint{1805.10138}.

\bibitem[{\citenamefont{Chiappetta et~al.}(1994)\citenamefont{Chiappetta,
  Colangelo, De~Felice, Nardulli, and Pasquariello}}]{Chiappetta:1993zv}
\bibinfo{author}{\bibfnamefont{P.}~\bibnamefont{Chiappetta}},
  \bibinfo{author}{\bibfnamefont{P.}~\bibnamefont{Colangelo}},
  \bibinfo{author}{\bibfnamefont{P.}~\bibnamefont{De~Felice}},
  \bibinfo{author}{\bibfnamefont{G.}~\bibnamefont{Nardulli}}, \bibnamefont{and}
  \bibinfo{author}{\bibfnamefont{G.}~\bibnamefont{Pasquariello}},
  \bibinfo{journal}{Phys. Lett.} \textbf{\bibinfo{volume}{B322}},
  \bibinfo{pages}{219} (\bibinfo{year}{1994}), \eprint{hep-ph/9401343}.

\bibitem[{\citenamefont{Cogan et~al.}(2015)\citenamefont{Cogan, Kagan, Strauss,
  and Schwarztman}}]{Cogan:2014oua}
\bibinfo{author}{\bibfnamefont{J.}~\bibnamefont{Cogan}},
  \bibinfo{author}{\bibfnamefont{M.}~\bibnamefont{Kagan}},
  \bibinfo{author}{\bibfnamefont{E.}~\bibnamefont{Strauss}}, \bibnamefont{and}
  \bibinfo{author}{\bibfnamefont{A.}~\bibnamefont{Schwarztman}},
  \bibinfo{journal}{JHEP} \textbf{\bibinfo{volume}{02}}, \bibinfo{pages}{118}
  (\bibinfo{year}{2015}), \eprint{1407.5675}.

\bibitem[{\citenamefont{de~Oliveira et~al.}(2016)\citenamefont{de~Oliveira,
  Kagan, Mackey, Nachman, and Schwartzman}}]{deOliveira:2015xxd}
\bibinfo{author}{\bibfnamefont{L.}~\bibnamefont{de~Oliveira}},
  \bibinfo{author}{\bibfnamefont{M.}~\bibnamefont{Kagan}},
  \bibinfo{author}{\bibfnamefont{L.}~\bibnamefont{Mackey}},
  \bibinfo{author}{\bibfnamefont{B.}~\bibnamefont{Nachman}}, \bibnamefont{and}
  \bibinfo{author}{\bibfnamefont{A.}~\bibnamefont{Schwartzman}},
  \bibinfo{journal}{JHEP} \textbf{\bibinfo{volume}{07}}, \bibinfo{pages}{069}
  (\bibinfo{year}{2016}), \eprint{1511.05190}.

\bibitem[{\citenamefont{Almeida et~al.}(2015)\citenamefont{Almeida, Backović,
  Cliche, Lee, and Perelstein}}]{Almeida:2015jua}
\bibinfo{author}{\bibfnamefont{L.~G.} \bibnamefont{Almeida}},
  \bibinfo{author}{\bibfnamefont{M.}~\bibnamefont{Backović}},
  \bibinfo{author}{\bibfnamefont{M.}~\bibnamefont{Cliche}},
  \bibinfo{author}{\bibfnamefont{S.~J.} \bibnamefont{Lee}}, \bibnamefont{and}
  \bibinfo{author}{\bibfnamefont{M.}~\bibnamefont{Perelstein}},
  \bibinfo{journal}{JHEP} \textbf{\bibinfo{volume}{07}}, \bibinfo{pages}{086}
  (\bibinfo{year}{2015}), \eprint{1501.05968}.

\bibitem[{\citenamefont{Pearkes et~al.}(2017)\citenamefont{Pearkes, Fedorko,
  Lister, and Gay}}]{Pearkes:2017hku}
\bibinfo{author}{\bibfnamefont{J.}~\bibnamefont{Pearkes}},
  \bibinfo{author}{\bibfnamefont{W.}~\bibnamefont{Fedorko}},
  \bibinfo{author}{\bibfnamefont{A.}~\bibnamefont{Lister}}, \bibnamefont{and}
  \bibinfo{author}{\bibfnamefont{C.}~\bibnamefont{Gay}} (\bibinfo{year}{2017}),
  \eprint{1704.02124}.

\bibitem[{\citenamefont{Baldi et~al.}(2016)\citenamefont{Baldi, Bauer, Eng,
  Sadowski, and Whiteson}}]{Baldi:2016fql}
\bibinfo{author}{\bibfnamefont{P.}~\bibnamefont{Baldi}},
  \bibinfo{author}{\bibfnamefont{K.}~\bibnamefont{Bauer}},
  \bibinfo{author}{\bibfnamefont{C.}~\bibnamefont{Eng}},
  \bibinfo{author}{\bibfnamefont{P.}~\bibnamefont{Sadowski}}, \bibnamefont{and}
  \bibinfo{author}{\bibfnamefont{D.}~\bibnamefont{Whiteson}},
  \bibinfo{journal}{Phys. Rev.} \textbf{\bibinfo{volume}{D93}},
  \bibinfo{pages}{094034} (\bibinfo{year}{2016}), \eprint{1603.09349}.

\bibitem[{\citenamefont{Guest et~al.}(2016)\citenamefont{Guest, Collado, Baldi,
  Hsu, Urban, and Whiteson}}]{Guest:2016iqz}
\bibinfo{author}{\bibfnamefont{D.}~\bibnamefont{Guest}},
  \bibinfo{author}{\bibfnamefont{J.}~\bibnamefont{Collado}},
  \bibinfo{author}{\bibfnamefont{P.}~\bibnamefont{Baldi}},
  \bibinfo{author}{\bibfnamefont{S.-C.} \bibnamefont{Hsu}},
  \bibinfo{author}{\bibfnamefont{G.}~\bibnamefont{Urban}}, \bibnamefont{and}
  \bibinfo{author}{\bibfnamefont{D.}~\bibnamefont{Whiteson}},
  \bibinfo{journal}{Phys. Rev.} \textbf{\bibinfo{volume}{D94}},
  \bibinfo{pages}{112002} (\bibinfo{year}{2016}), \eprint{1607.08633}.

\bibitem[{\citenamefont{Lonnblad et~al.}(1990)\citenamefont{Lonnblad, Peterson,
  and Rognvaldsson}}]{Lonnblad:1990bi}
\bibinfo{author}{\bibfnamefont{L.}~\bibnamefont{Lonnblad}},
  \bibinfo{author}{\bibfnamefont{C.}~\bibnamefont{Peterson}}, \bibnamefont{and}
  \bibinfo{author}{\bibfnamefont{T.}~\bibnamefont{Rognvaldsson}},
  \bibinfo{journal}{Phys. Rev. Lett.} \textbf{\bibinfo{volume}{65}},
  \bibinfo{pages}{1321} (\bibinfo{year}{1990}).

\bibitem[{\citenamefont{Lonnblad et~al.}(1991)\citenamefont{Lonnblad, Peterson,
  and Rognvaldsson}}]{Lonnblad:1990qp}
\bibinfo{author}{\bibfnamefont{L.}~\bibnamefont{Lonnblad}},
  \bibinfo{author}{\bibfnamefont{C.}~\bibnamefont{Peterson}}, \bibnamefont{and}
  \bibinfo{author}{\bibfnamefont{T.}~\bibnamefont{Rognvaldsson}},
  \bibinfo{journal}{Nucl. Phys.} \textbf{\bibinfo{volume}{B349}},
  \bibinfo{pages}{675} (\bibinfo{year}{1991}).

\bibitem[{\citenamefont{Peterson et~al.}(1994)\citenamefont{Peterson,
  Rognvaldsson, and Lonnblad}}]{Peterson:1993nk}
\bibinfo{author}{\bibfnamefont{C.}~\bibnamefont{Peterson}},
  \bibinfo{author}{\bibfnamefont{T.}~\bibnamefont{Rognvaldsson}},
  \bibnamefont{and} \bibinfo{author}{\bibfnamefont{L.}~\bibnamefont{Lonnblad}},
  \bibinfo{journal}{Comput. Phys. Commun.} \textbf{\bibinfo{volume}{81}},
  \bibinfo{pages}{185} (\bibinfo{year}{1994}).

\bibitem[{\citenamefont{Komiske et~al.}(2017)\citenamefont{Komiske, Metodiev,
  and Schwartz}}]{Komiske:2016rsd}
\bibinfo{author}{\bibfnamefont{P.~T.} \bibnamefont{Komiske}},
  \bibinfo{author}{\bibfnamefont{E.~M.} \bibnamefont{Metodiev}},
  \bibnamefont{and} \bibinfo{author}{\bibfnamefont{M.~D.}
  \bibnamefont{Schwartz}}, \bibinfo{journal}{JHEP}
  \textbf{\bibinfo{volume}{01}}, \bibinfo{pages}{110} (\bibinfo{year}{2017}),
  \eprint{1612.01551}.

\bibitem[{\citenamefont{Lecun et~al.}(1989)\citenamefont{Lecun, Boser, Denker,
  Henderson, Howard, Hubbard, and Jackel}}]{3a592b7a61c24ba7a7a35bf0983c9b68}
\bibinfo{author}{\bibfnamefont{Y.}~\bibnamefont{Lecun}},
  \bibinfo{author}{\bibfnamefont{B.}~\bibnamefont{Boser}},
  \bibinfo{author}{\bibfnamefont{J.}~\bibnamefont{Denker}},
  \bibinfo{author}{\bibfnamefont{D.}~\bibnamefont{Henderson}},
  \bibinfo{author}{\bibfnamefont{R.}~\bibnamefont{Howard}},
  \bibinfo{author}{\bibfnamefont{W.}~\bibnamefont{Hubbard}}, \bibnamefont{and}
  \bibinfo{author}{\bibfnamefont{L.}~\bibnamefont{Jackel}},
  \bibinfo{journal}{Neural computation} \textbf{\bibinfo{volume}{1}},
  \bibinfo{pages}{541} (\bibinfo{year}{1989}), ISSN \bibinfo{issn}{0899-7667}.

\bibitem[{\citenamefont{Srivastava et~al.}(2014)\citenamefont{Srivastava,
  Hinton, Krizhevsky, Sutskever, and Salakhutdinov}}]{JMLR:v15:srivastava14a}
\bibinfo{author}{\bibfnamefont{N.}~\bibnamefont{Srivastava}},
  \bibinfo{author}{\bibfnamefont{G.}~\bibnamefont{Hinton}},
  \bibinfo{author}{\bibfnamefont{A.}~\bibnamefont{Krizhevsky}},
  \bibinfo{author}{\bibfnamefont{I.}~\bibnamefont{Sutskever}},
  \bibnamefont{and}
  \bibinfo{author}{\bibfnamefont{R.}~\bibnamefont{Salakhutdinov}},
  \bibinfo{journal}{Journal of Machine Learning Research}
  \textbf{\bibinfo{volume}{15}}, \bibinfo{pages}{1929} (\bibinfo{year}{2014}),
  \urlprefix\url{http://jmlr.org/papers/v15/srivastava14a.html}.

\bibitem[{\citenamefont{Chollet et~al.}(2015)}]{chollet2015keras}
\bibinfo{author}{\bibfnamefont{F.}~\bibnamefont{Chollet}} \bibnamefont{et~al.},
  \emph{\bibinfo{title}{Keras}},
  \bibinfo{howpublished}{\url{https://github.com/keras-team/keras}}
  (\bibinfo{year}{2015}).

\bibitem[{\citenamefont{Yan et~al.}(2016)\citenamefont{Yan, Watanuki, Fujii,
  Ishikawa, Jeans, Strube, Tian, and Yamamoto}}]{Yan:2016xyx}
\bibinfo{author}{\bibfnamefont{J.}~\bibnamefont{Yan}},
  \bibinfo{author}{\bibfnamefont{S.}~\bibnamefont{Watanuki}},
  \bibinfo{author}{\bibfnamefont{K.}~\bibnamefont{Fujii}},
  \bibinfo{author}{\bibfnamefont{A.}~\bibnamefont{Ishikawa}},
  \bibinfo{author}{\bibfnamefont{D.}~\bibnamefont{Jeans}},
  \bibinfo{author}{\bibfnamefont{J.}~\bibnamefont{Strube}},
  \bibinfo{author}{\bibfnamefont{J.}~\bibnamefont{Tian}}, \bibnamefont{and}
  \bibinfo{author}{\bibfnamefont{H.}~\bibnamefont{Yamamoto}},
  \bibinfo{journal}{Phys. Rev.} \textbf{\bibinfo{volume}{D94}},
  \bibinfo{pages}{113002} (\bibinfo{year}{2016}), \eprint{1604.07524}.

\bibitem[{\citenamefont{Xin~Mo}(2017)}]{XinMo}
\bibinfo{author}{\bibfnamefont{G.~L.} \bibnamefont{Xin~Mo}}
  (\bibinfo{collaboration}{CEPC Working Group}),
  \emph{\bibinfo{title}{{Generated Sample Stauts for CEPC Simulation Studies
  (CEPC Note in preparation)}}} (\bibinfo{year}{2017}).

\bibitem[{\citenamefont{Kilian et~al.}(2011)\citenamefont{Kilian, Ohl, and
  Reuter}}]{Kilian:2007gr}
\bibinfo{author}{\bibfnamefont{W.}~\bibnamefont{Kilian}},
  \bibinfo{author}{\bibfnamefont{T.}~\bibnamefont{Ohl}}, \bibnamefont{and}
  \bibinfo{author}{\bibfnamefont{J.}~\bibnamefont{Reuter}},
  \bibinfo{journal}{Eur. Phys. J.} \textbf{\bibinfo{volume}{C71}},
  \bibinfo{pages}{1742} (\bibinfo{year}{2011}), \eprint{0708.4233}.

\bibitem[{\citenamefont{Moretti et~al.}(2001)\citenamefont{Moretti, Ohl, and
  Reuter}}]{Moretti:2001zz}
\bibinfo{author}{\bibfnamefont{M.}~\bibnamefont{Moretti}},
  \bibinfo{author}{\bibfnamefont{T.}~\bibnamefont{Ohl}}, \bibnamefont{and}
  \bibinfo{author}{\bibfnamefont{J.}~\bibnamefont{Reuter}}, pp.
  \bibinfo{pages}{1981--2009} (\bibinfo{year}{2001}), \eprint{hep-ph/0102195}.

\bibitem[{\citenamefont{Sjostrand et~al.}(2006)\citenamefont{Sjostrand, Mrenna,
  and Skands}}]{Sjostrand:2006za}
\bibinfo{author}{\bibfnamefont{T.}~\bibnamefont{Sjostrand}},
  \bibinfo{author}{\bibfnamefont{S.}~\bibnamefont{Mrenna}}, \bibnamefont{and}
  \bibinfo{author}{\bibfnamefont{P.~Z.} \bibnamefont{Skands}},
  \bibinfo{journal}{JHEP} \textbf{\bibinfo{volume}{05}}, \bibinfo{pages}{026}
  (\bibinfo{year}{2006}), \eprint{hep-ph/0603175}.

\bibitem[{\citenamefont{Bellm et~al.}(2016)}]{Bellm:2015jjp}
\bibinfo{author}{\bibfnamefont{J.}~\bibnamefont{Bellm}} \bibnamefont{et~al.},
  \bibinfo{journal}{Eur. Phys. J.} \textbf{\bibinfo{volume}{C76}},
  \bibinfo{pages}{196} (\bibinfo{year}{2016}), \eprint{1512.01178}.

\bibitem[{\citenamefont{Cacciari et~al.}(2012)\citenamefont{Cacciari, Salam,
  and Soyez}}]{Cacciari:2011ma}
\bibinfo{author}{\bibfnamefont{M.}~\bibnamefont{Cacciari}},
  \bibinfo{author}{\bibfnamefont{G.~P.} \bibnamefont{Salam}}, \bibnamefont{and}
  \bibinfo{author}{\bibfnamefont{G.}~\bibnamefont{Soyez}},
  \bibinfo{journal}{Eur. Phys. J.} \textbf{\bibinfo{volume}{C72}},
  \bibinfo{pages}{1896} (\bibinfo{year}{2012}).

\bibitem[{\citenamefont{Simonyan and Zisserman}(2014)}]{Simonyan14c}
\bibinfo{author}{\bibfnamefont{K.}~\bibnamefont{Simonyan}} \bibnamefont{and}
  \bibinfo{author}{\bibfnamefont{A.}~\bibnamefont{Zisserman}},
  \bibinfo{journal}{CoRR} \textbf{\bibinfo{volume}{abs/1409.1556}}
  (\bibinfo{year}{2014}).

\bibitem[{\citenamefont{Kingma and Ba}(2014)}]{Kingma:2014vow}
\bibinfo{author}{\bibfnamefont{D.~P.} \bibnamefont{Kingma}} \bibnamefont{and}
  \bibinfo{author}{\bibfnamefont{J.}~\bibnamefont{Ba}} (\bibinfo{year}{2014}),
  \eprint{1412.6980}.

\bibitem[{\citenamefont{Krizhevsky et~al.}(2012)\citenamefont{Krizhevsky,
  Sutskever, and Hinton}}]{NIPS2012_4824}
\bibinfo{author}{\bibfnamefont{A.}~\bibnamefont{Krizhevsky}},
  \bibinfo{author}{\bibfnamefont{I.}~\bibnamefont{Sutskever}},
  \bibnamefont{and} \bibinfo{author}{\bibfnamefont{G.~E.}
  \bibnamefont{Hinton}}, in \emph{\bibinfo{booktitle}{Advances in Neural
  Information Processing Systems 25}}, edited by
  \bibinfo{editor}{\bibfnamefont{F.}~\bibnamefont{Pereira}},
  \bibinfo{editor}{\bibfnamefont{C.~J.~C.} \bibnamefont{Burges}},
  \bibinfo{editor}{\bibfnamefont{L.}~\bibnamefont{Bottou}}, \bibnamefont{and}
  \bibinfo{editor}{\bibfnamefont{K.~Q.} \bibnamefont{Weinberger}}
  (\bibinfo{publisher}{Curran Associates, Inc.}, \bibinfo{year}{2012}), pp.
  \bibinfo{pages}{1097--1105}.

\bibitem[{\citenamefont{Guo et~al.}(2018)\citenamefont{Guo, Li, Li, Xu, and
  Zhang}}]{Guo:2018hbv}
\bibinfo{author}{\bibfnamefont{J.}~\bibnamefont{Guo}},
  \bibinfo{author}{\bibfnamefont{J.}~\bibnamefont{Li}},
  \bibinfo{author}{\bibfnamefont{T.}~\bibnamefont{Li}},
  \bibinfo{author}{\bibfnamefont{F.}~\bibnamefont{Xu}}, \bibnamefont{and}
  \bibinfo{author}{\bibfnamefont{W.}~\bibnamefont{Zhang}},
  \bibinfo{journal}{Phys. Rev.} \textbf{\bibinfo{volume}{D98}},
  \bibinfo{pages}{076017} (\bibinfo{year}{2018}), \eprint{1805.10730}.

\end{thebibliography}

\end{document}